\documentstyle[12pt,equations,cite]{article}
\newcommand{\smallz}{{\scriptscriptstyle Z}} 
\newcommand{\smallw}{{\scriptscriptstyle W}} %
\newcommand{\smallh}{{\scriptscriptstyle H}} %

\newcommand{\acur}{ {\hat \alpha} }
\newcommand{\as}{ {\hat \alpha}_s }
\newcommand{\mz}{M_\smallz}
\newcommand{\mw}{M_\smallw}
\newcommand{\mh}{M_\smallh}
\newcommand{\mt}{M_t}
\newcommand{\mut}{\mu_t}

\newcommand{\seff}{\mbox{$\sin^2 \theta_{{ eff}}^{lept}$} }
 
\newcommand{\scur}{\mbox{$\hat{s}^2$}}
\newcommand{\sincur}{\mbox{$\sin^{2}\!\hat{\theta}_{\scriptscriptstyle W} 
                           (\mz)$}}
\newcommand{\ccur}{\mbox{$\hat{c}^2$}}

\newcommand{\dr}{\mbox{$ \Delta r$}}

\newcommand{\drcar}{\mbox{$\Delta \hat{r}$}}

\newcommand{\azz}{\mbox{$ A_{\smallz \smallz} $}}
\newcommand{\aww}{\mbox{$ A_{\smallw \smallw} $}}

\newcommand{\gd}{\mbox{$ O(g^4 \mt^2/ \mw^2)  $}}

\newcommand{\gq}{\mbox{$ O(g^4 \mt^4/\mw^4) $}}
\newcommand{\ew}{electroweak}
\newcommand{\msbar}{\overline{\rm MS}}
\def\lequiv{\raise 0.4ex \hbox{$<$} \kern -0.8 em \lower 0.62 ex \hbox{$\sim$}}
\def\gequiv{\raise 0.4ex \hbox{$>$} \kern -0.7 em \lower 0.62 ex \hbox{$\sim$}}
\newcommand{\equ}[1]{Eq.\,(\ref{#1})}

\newcommand{\Eqs}[2]{Eqs.\,(\ref{#1}) and (\ref{#2})}
\newcommand{\efe}[1]{Ref.\cite{#1}}
\newcommand{\efs}[2]{Refs.\cite{#1,#2}}
\newcommand{\be}{\begin{equation}}
\newcommand{\ee}{\end{equation}}
\newcommand{\een}{\end{subequations}}
\newcommand{\ben}{\begin{subequations}}
\newcommand{\beq}{\begin{eqalignno}}
\newcommand{\eeq}{\end{eqalignno}}
\newcommand{\bea}{\begin{eqnarray}}
\newcommand{\eea}{\end{eqnarray}}
\newenvironment{appendletter}
 {
  \typeout{ Starting Appendix \thesection }
  \setcounter{equation}{0}
  
 }{
  \typeout{Appendix done}
 }
\renewcommand{\thefootnote}{\fnsymbol{footnote} }
\begin{document}
\begin{titlepage}

\begin{flushright}
        \small
        NYU-TH-96/11/02\\
        MPI-PhT/96-118
\end{flushright}
\vspace{1.2cm} 
\begin{center}
{\Large\bf 
Precise calculation of   $\mw$, $\sincur$, and $\seff$}
 \\
\vspace{2cm}
{\sc Giuseppe Degrassi$^a$\footnote{
On leave from  Dipartimento di Fisica, Universit\`a  di Padova,
Padua, Italy.},
 Paolo Gambino$^a$, and Alberto Sirlin$^b$}\\
\vspace{.4cm}
{\em $^a$ Max Planck Institut f\"ur Physik, Werner Heisenberg Institut,\\
 F\"ohringer Ring 6, D80805 M\"unchen, Germany}\\
\vspace{.3cm}
{\em $^b$  Department of Physics, New York University\\
4 Washington Place, New York, NY 10003, USA}
\end{center}
\vspace{2cm} 
\begin{center}
{\bf Abstract}
\end{center}
\vspace{0.2cm}
The two-loop $\gd$  corrections 
are incorporated in the theoretical calculation of $\mw$, $\sincur$, and 
$\seff$, as functions of $\mh$. The analysis is carried out in a 
previously proposed 
$\msbar$ formulation and two novel on-shell resummation schemes. It is
found that the inclusion of the new effects sharply decreases
the scheme and residual scale dependence of the calculations. QCD
corrections are incorporated in two different approaches.
 Comparison with the world average of $\seff$ leads to
$\mh= 127^{+143}_{-71}$ GeV and $\mw= 80.367\pm0.048$GeV, 
with small variations among the six calculations.
\vfill\nopagebreak
\end{titlepage}
\setcounter{footnote}{0}
\renewcommand{\thefootnote}{\arabic{footnote}}

The theoretical calculation of the $W$-boson mass $\mw$,
the $\msbar$ parameter \sincur, and the effective mixing parameter \seff\
represents one of the most important applications of the Standard Model (SM) 
at the level of its quantum corrections.
We recall that \seff\ is determined from on-resonance observables at LEP and 
SLC, while \sincur\ is theoretically very important, particularly in the 
context of GUT studies. The aim of this paper is to present an accurate
calculation of  these parameters that, in the case of $\mw$ and \sincur,
includes the recently evaluated two-loop contributions of  
\gd\ \cite{physlett}. The corresponding contributions for the evaluation of
\seff are provided in the present paper. Although nominally subleading 
relative to the two-loop terms of \gq, these effects have been found to 
be of comparable magnitude \cite{physlett}.
Our strategy is to study the incorporation of the \gd\ contributions in 
different renormalization schemes, and to investigate the residual scale 
dependence.
In fact, as it will be shown, 
the inclusion of the new results greatly decreases 
the scheme and residual scale dependence of important radiative corrections.
QCD corrections are evaluated following two different approaches.

We recall that, using the accurately known parameters $\alpha$, $G_\mu$, and 
$\mz$ as inputs, $\mw$ can be calculated from the expression \cite{si80}
\be
s^2 c^2 = A^2/\mz^2 (1- \Delta r),\label{uno}
\ee
where $s^2=1-c^2 $ is an abbreviation for $\sin^2 \theta_\smallw\equiv
1-\mw^2/\mz^2$ ($\mw$ and $\mz$ are the physical masses of the intermediate 
bosons), $A^2=\pi \alpha/\sqrt{2} G_\mu = (37.2802\, {\rm GeV})^2$,
and $ \Delta r$ is the relevant radiative correction \cite{si80}.
In order to incorporate the \gd\ effects, it is necessary to 
specify the renormalization scheme employed in the evaluation of the 
one-loop contributions. In \efe{physlett} this was done in the $\msbar$ 
framework of \efe{DFS}, where
\bea
&&1- \Delta r =  \left[1-\drcar_\smallw\right] \,\scur/s^2,\label{due}\\
&& \scur/s^2  =  1 + (c^2/s^2) (\hat{e}^2/\scur) 
\ \Delta\hat{\rho},\label{tre}\\
&&\Delta\hat{\rho}= 
 {\rm Re}\,\left[ \aww(\mw^2)- \ccur \azz(\mz^2)\right]_{\msbar}/\mw^2,
\label{quattro}\\
&&\drcar_\smallw = - (2\delta e/e)_{\msbar} + 
(e^2/\scur) \hat{f},\label{cinque}\\
&& \hat{f} =
\left[({\rm{Re}}
 \aww(\mw^2) - \aww(0))/\mw^2 + V_\smallw + \mw^2 B_\smallw
\right]_{\msbar}.\label{sei}
\eea
In these expressions $\msbar$ denotes both the $\msbar$ renormalization (i.e.
 the pole subtraction) and the choice $\mu=\mz$ for the 't-Hooft scale, $\scur
=1-\ccur$ is an abbreviation for \sincur,
$2\delta e/e$ is the charge renormalization counterterm, 
$\hat{e}^2= e^2/(1+2\delta e/e)_{\msbar}$ the $\msbar$ electromagnetic 
coupling at $\mu=\mz$, and $\aww(q^2)$ and $\azz(q^2)$ the transverse $W$ and
 $Z^0$ self-energies \cite{si80,DFS,ms80} with  $\hat{g}^2= \hat{e}^2/\scur$ 
factored out.
In this paper we also factor out 
$\hat{g}^2$ in the definition of $\Delta\hat{\rho}$.
As $\mt$ has been measured, it is simplest to use a pure $\msbar$
subtraction, without decoupling the top contribution 
to $\delta e/e$ and the  mixing self-energy
 $A_{\gamma\smallz}(\mz^2)$ \cite{FKS}.
The term $V_\smallw + \mw^2 B_\smallw$ in \equ{sei}
represents vertex and box diagram contributions to $\mu$ decay, modulo
a factor $e^2/\scur$ \cite{si80}.
QCD corrections to the self-energies are given, for example,
in \efe{FKS,cks} and are updated  in this paper.

As shown in \efe{physlett},
 in the $\msbar$ framework of Eqs. (\ref{uno}-\ref{sei})
the \gd\ corrections are incorporated by adding two-loop
irreducible contributions, $\drcar_\smallw^{(2)}$
and $(\hat{e}^2/\scur)\Delta\hat{\rho}^{(2)}$, to the one-loop expressions 
$\drcar_\smallw^{(1)}$
and $(\hat{e}^2/\scur)\Delta\hat{\rho}^{(1)}$, respectively.
The correction $(\hat{e}^2/\scur)\Delta\hat{\rho}^{(2)}$ also includes the 
previously evaluated irreducible contributions of \gq\ \cite{barb}.

In order to discuss the scheme dependence, we note that a resummation 
formula analogous to \equ{due} can be obtained in the on-shell (OS)
renormalization
scheme of \efe{si80}.
For brevity, we show 
how it follows from  Eqs.(\ref{uno}-\ref{sei}). Combining 
Eqs. (\ref{due},\ref{tre},\ref{cinque}) we have
\be
1-\dr=  1+\, ({2\delta e/e})_{\msbar} \,
+ \,(c^2/s^2)\ (e^2/\scur)\ \Delta\hat{\rho}
\,-\,(e^2/s^2) \hat{f}.\label{sette}
\ee
In order to express $e^2/\scur$ in terms of $G_\mu$, 
we note that Eqs. (\ref{uno}-\ref{tre}) lead to
\be
e^2/\scur = (G_\mu/\sqrt{2}) \, 8 \,\mw^2 \left[
1 + (2\delta e/e)_{\msbar} -(e^2/\scur) \hat{f}
\right].\label{otto}
\ee
Approximating $e^2/\scur\to e^2/s^2$ in the square bracket and inserting 
\equ{otto} in the $\Delta\hat{\rho}$ term of \equ{sette}, we have
\be
1-\dr=
\left[ 1+ ({2\delta e/e})_{\msbar} -
\frac{e^2}{s^2} \hat{f}\right]
\left(1+ \frac{c^2}{s^2} \,8 \mw^2 \frac{G_\mu}{\sqrt{2}}
\Delta\hat{\rho} \right).\label{nove}
\ee
The leading $\mt$ contribution to $\Delta\hat{\rho}^{(1)}$ is
$3\mt^2/(64\pi^2\mw^2)$, which is independent of $\scur$.
At the one-loop level $({2\delta e/e})_{\msbar}$
is also independent of $\scur $ and $\hat{e}$ \cite{DFS,FKS},
 and $(e^2/s^2) \hat{f}^{(1)}$ does not involve terms proportional 
to $\mt^2$ or large logarithms $\ln(\mz/m_f)$, where
$m_f$ is a small fermion mass . However,
non-leading one-loop contributions contained in $(e^2/s^2) \hat{f}^{(1)}$
and $(G_\mu/\sqrt{2})\,8\mw^2 \Delta\hat{\rho}^{(1)}$ do depend on $\scur$.
To obtain a resummation formula involving only on-shell parameters,
we replace 
$\scur= s^2\, [1 + (c^2/s^2)\, (\hat{e}^2/\scur)\,\Delta\hat{\rho}]$ 
in $\hat{f}(\scur)$ and $\Delta\hat{\rho}(\scur)$.
Calling $\bar{f}(s^2)$ and $\Delta\bar{\rho}(s^2)$ the resulting functions,
we obtain
\be
1-\dr= \left( 1+ \left.\frac{2\delta e}{e}\right|_{\msbar} -
\frac{e^2}{s^2} \bar{f}(s^2)\right)
\left(1+ \frac{c^2}{s^2} \frac{8 \mw^2 G_\mu}{\sqrt{2}}
\Delta\bar{\rho}(s^2) \right)\label{dieci}
\ee
which is an on-shell counterpart of the $\msbar$ expression of \equ{due}.
We note from \equ{tre} that $\scur-s^2= c^2 (\hat{e}^2/\scur)
\Delta\hat{\rho}\approx 3 \,c^2\, x_t + ...$
where
$
x_t= G_\mu \mt^2/8\pi^2 \sqrt{2} 
$
and the ellipses represent subleading  contributions. Thus the 
replacement
$\scur= s^2\, [1 + (c^2/s^2)\,3\,x_t+ ...]$ in $\hat{f}^{(1)}(\scur)$ and 
$\Delta\hat{\rho}^{(1)}(\scur)$ induces additional contributions of 
\gd.
The corresponding functions, called $\bar{f}^{(2)}_{add}$
and $\Delta\bar{\rho}^{(2)}_{add}$, respectively, are given in the Appendix.
Therefore in the OS scheme we obtain
\bea
\bar{f}(s^2)&=& \hat{f}^{(1)}(s^2) + 
\bar{f}^{(2)}(s^2) 
\label{dodici}\\
\Delta\bar{\rho}(s^2)&=& \Delta\hat{\rho}^{(1)}(s^2) + \Delta\bar{\rho}^{(2)}
(s^2)
\label{tredici},
\eea
where
 $\bar{f}^{(2)}(s^2)= \hat{f}^{(2)}(s^2)+ \bar{f}^{(2)}_{add}(s^2)$,
$\Delta\bar{\rho}^{(2)}(s^2)=\Delta\hat{\rho}^{(2)}(s^2)+
\Delta\bar{\rho}^{(2)}_{add}(s^2)$.
The amplitude $(e^2/s^2) \hat{f}^{(2)}$ is given by Eqs.(7a,b) of 
\efe{physlett} multiplied by $(\alpha/\pi s^2) N_c\, x_t$ ($N_c=3$),
while $(G_\mu/\sqrt{2})\, 8 \mw^2 \, \Delta\hat{\rho}^{(2)}$ is given 
by Eqs.(10a,b) of the same reference
multiplied by $N_c x_t^2$. 

An alternative OS resummation can be obtained by combining
$(e^2/\scur)\, \Delta\hat{\rho}$ in \equ{sette} with \equ{tre}
and once more replacing $\hat{f}\to\bar{f}$,
$\Delta\hat{\rho}\to\Delta\bar{\rho}$.
This leads to
\be
\dr= \dr^{(1)} + \dr^{(2)} + \left(\frac{c^2}{s^2} G_\mu \frac{
8\mw^2}{\sqrt{2}}\Delta\bar{\rho}(s^2)\right)^2 (1-\Delta\alpha),
\label{quattordici}
\ee
where $\dr^{(1)}$ is the original one-loop OS result of \efs{si80}{ms80},
expressed in terms of $\alpha$ and $s^2$, $\dr^{(2)}=
(e^2/s^2)\bar{f}^{(2)} - (c^2/s^2)(e^2/s^2)\Delta\bar{\rho}^{(2)}$,
and $\Delta\alpha$ is 
the renormalized photon
vacuum polarization function at $q^2=\mz^2$. If in the last term we only 
retain the two-loop contributions proportional to $\mt^2$ and $\mt^4$,
\equ{quattordici} reduces to
\be
\Delta r = \Delta r^{(1)}
 + \Delta r^{(2)}  
  + \left( \frac{c^2}{s^2}\right)^2 
N_c\, x_t \left( 2\frac{e^2}{s^2} \Delta\bar{\rho}^{(1)} -
N_c\frac{\alpha}{16\pi \,s^2} \frac{\mt^2}{\mw^2}\right)
\label{quindici}.
\ee
The last terms in Eq.\,(\ref{quattordici},\ref{quindici})
represent higher order reducible contributions induced by resummation of 
one-loop corrections, while $\dr^{(2)}$ contains the corresponding irreducible 
components. Eqs.\,(\ref{due},\ref{dieci},\ref{quattordici},\ref{quindici})
satisfy the important property that, when inserted in \equ{uno},
every factor of $\alpha$ in that equation is matched by a factor 
$(1+ 2\delta e/e)_{\msbar}^{-1}$ or, equivalently, $(1-\Delta\alpha)^{-1}$
\cite{si84}.

In the previous considerations we have set $\mu=\mz$.
However, one may also consider the case of a general $\mu$.
Although the exact $\dr$, being a physical observable, is $\mu$-independent,
we note that the resummation formulae of Eqs. 
(\ref{due},\ref{dieci},\ref{quattordici})
involve terms quadratic in gauge-independent, but 
$\mu$-dependent one-loop amplitudes.  As a consequence, 
these resummation formulae contain a $\mu$-dependence in $O(g^4)$.
We have verified that the inclusion of the irreducible two-loop effects,
$\hat{f}^{(2)}$ and $\Delta\hat{\rho}^{(2)}$ in \equ{due},
and $\bar{f}^{(2)}$ and $\Delta\bar{\rho}^{(2)}$ in 
Eqs.(\ref{dieci},\ref{quattordici}),
cancels the $\mu$-dependence through \gd,
the order of validity of the calculation. There remains a significantly 
smaller  $\mu$-dependence of $O(g^4)$ without $\mt^2$ enhancement
factors due to the fact that complete 
$O(g^4)$ corrections have not yet been evaluated. 
 On the other hand, \equ{quindici} is exactly $\mu$-independent
since it retains only the complete two-loop contributions proportional to 
$\mt^4$ and $\mt^2$.
We refer to Eqs.(\ref{due},\ref{tre}), \equ{dieci}, and \equ{quindici}
as the $\msbar$, OSI,
and OSII schemes, respectively.
 Their numerical difference will give us a measure of the scheme dependence
of $\dr$ and the corresponding $\mw$ predictions.

The effective parameter \seff\ is obtained from $\scur$ or $s^2$ by means 
of the relations
\be
\sin^2\theta_{eff}^{lept}\ =\ \hat{k}(\mz^2)\  \sincur\ = \ k(\mz^2) \  s^2
\label{sedici}
\ee
where $\hat{k}(q^2)$ and ${k}(q^2)$ are the real parts of \ew\
form factors evaluated at $q^2=\mz^2$, and $\scur$ and $s^2$ are related
 by \equ{tre}.
The amplitude $\hat{k}$, evaluated in the $\msbar$ scheme with $\mu=\mz$,
can be expressed as 
$\hat{k}= 1 + (\hat{e}^2/\scur)\, (\Delta \hat{k}^{(1)} 
+ \Delta \hat{k}^{(2)})$. The one-loop contribution 
$\Delta \hat{k}^{(1)}$ is given in \efe{rel},
 with the understanding that
the top contribution to the mixing self-energy
 $A_{\gamma\smallz}(\mz^2)$ is not decoupled and, 
following the discussion of that work,
certain two-loop effects induced by the imaginary parts of the self-energies
are included.
We recall that, for large $\mt$, $\Delta \hat{k}^{(1)} $ grows like 
$\ln(\mt^2/\mz^2)$.
The \gd\ corrections are contained in $(\hat{e}^2/\scur)\,\Delta 
\hat{k}^{(2)}$, which is given in \equ{app1}, multiplied by 
$N_c (\acur/4\pi\scur)^2\, \mt^2/(\ccur\mz^2)$. The on-shell
amplitude $k$ is obtained from $\hat{k}$ using \equ{tre},
$\hat{e}^2/\scur = (G_\mu/\sqrt{2}) \, 8 \,\mw^2 \left[
1  -(\hat{e}^2/\scur) \hat{f}\right]$
(which follows from \equ{otto}), approximating $(\hat{e}^2/\scur) \hat{f}(
\scur) \approx  (G_\mu/\sqrt{2}) \, 8 \,\mw^2 \bar{f}(s^2) $
 in the square bracket,
expressing $\scur$ in $\Delta \hat{k}^{(1)}(\scur)+\Delta \hat{k}^{(2)}
(\scur) $ in terms of $s^2$ and taking into account the additional \gd\
contribution $\Delta \bar{k}^{(2)}_{add}$ induced by the latter 
shift. Neglecting two-loop terms without $\mt^2$ enhancement
factors, this leads to
\be
k= 
 \left(1+\frac{8\mw^2 G_\mu}{\sqrt{2}} \Delta\bar{k}(s^2) \right)
\left[ 1+ \frac{c^2}{s^2} \frac{8\mw^2 G_\mu}{\sqrt{2}}\left(1-
\frac{8\mw^2 G_\mu}{\sqrt{2}}
\bar{f}^{(1)}(s^2)\right)
 \Delta\bar{\rho}(s^2) \right],\label{diciassette}
\ee
where $\Delta\bar{\rho}$ is defined in \equ{tredici} and $\Delta\bar{k}(s^2)=
\Delta\hat{k}^{(1)}(s^2)+ \Delta\bar{k}^{(2)}(s^2)$,
$\Delta\bar{k}^{(2)}(s^2)= \Delta \hat{k}^{(2)}(s^2) + \Delta 
\bar{k}^{(2)}_{add}(s^2)$. If we only retain two-loop effects with 
$\mt^4$ and $\mt^2$ enhancement factors, \equ{diciassette} reduces to
\be
k= 1 + \frac{8\mw^2 G_\mu}{\sqrt{2}} \left[ \Delta \bar{k}(s^2)
 + \frac{c^2}{s^2}
\Delta\bar{\rho}(s^2)  +  \frac{c^2}{s^2} \,N_c\, x_t \,\left(
 \Delta \bar{k}^{(1)}(s^2) 
-  \bar{f}^{(1)}(s^2) \right)\right],
\label{diciotto}
\ee
which is exactly $\mu$-independent.

We briefly explain the strategies 
followed in the evaluation of the QCD corrections.
When \ew\ amplitudes proportional to the 
squared top-quark mass are expressed in terms of
the pole mass $\mt$, the coefficients of $a^n(t)$ ($a\equiv\hat{\alpha}_s/\pi$,
$n=1,2$) are quite large, a feature that is absent when they are expressed
in terms of the running $\msbar$ mass  $\mut= \hat{m}_t
(\mut)$. Recent studies of the scale and scheme dependence of the leading 
$\mt$-dependent \ew\ corrections also suggest that the unknown QCD corrections
to the \gq\ contributions are significant when $\mt$ is employed \cite{ks95}.
These observations suggest the strategy of parametrizing the 
\ew\ amplitudes in terms of $\mut$. 
 We illustrate this approach in the most sensitive amplitude, 
namely the contributions to $\Delta\bar{\rho}$
and $\Delta\hat{\rho}$ proportional to powers of $\mt$.
 In the case of $\Delta\bar{\rho}$ we have
\be
\frac{8\mw^2 G_\mu}{\sqrt{2}} 
\Delta\bar{\rho}= N_c\, \frac{ G_\mu}{\sqrt{2}} \frac{\mut^2 }{8\pi^2}
\left[ 1 + \delta_{QCD}^{\msbar} + \frac{ G_\mu \mut^2} {8\sqrt{2}\pi^2}
R(r_\smallh, r_\smallz) +  \frac{ G_\mu \mw^2} {2\sqrt{2}\pi^2}
R_{add} \right]+...,
\label{diciannove}
\ee
where $R_{add}= 16 \pi^2 \Delta\bar{\rho}_{add} /(N_c\, x_t)$ and
\be
\delta_{QCD}^{\msbar}= -0.19325\, a(\mut) - 3.970\, a^2(\mut),
\label{venti}
\ee
$r_\smallh\equiv \mh^2/\mut^2$, $ r_\smallz\equiv\mz^2/\mut^2$,
$R(r_\smallh, r_\smallz) $ is the function given in Eqs.(10a,b)
of \efe{physlett} with $\mt\to\mut$ and $\ccur\to c^2$, and the ellipses 
stand for contributions not proportional to powers of $\mut$.
The coefficients in \equ{venti} are  much smaller than in 
 \cite{cks}
\be
\delta_{QCD}= -2.8599 \,a(\mt) - 14.594\, a^2(\mt),
\label{ventibis}
\ee
the QCD correction when $\Delta\bar{\rho}$ is expressed in terms of $\mt$.
In the case of $\Delta\hat{\rho}$ we have an analogous formula to 
\equ{diciannove}, except that one retains the $\scur $ dependence of $R$,
$R_{add}$ is absent and the couplings are expressed in terms of $\hat{e}^2/
\scur$. This strategy is  extended to the other \ew\
amplitudes using the perturbative relation
$\mt=[1+(4/3) \, a(\mut) + 8.236\,a^2(\mut)]\,\mut$ in the terms
of $O(\alpha)$ and $O(\alpha\as)$ of \efe{FKS}.
As $\mt$ is important in the interpretation of the experiments,
we employ 
\be
\mu_t/\mt=
[ 1+ (8/3) a^2(\hat{\mut}) - 4.47 a^3(\hat{\mut})]/[
1+ (4/3) a(\mut^*) - 1.072 a^2(\mut^*)],
\label{ventuno}
\ee
where $\hat{\mut}= 0.252\mt$, $\mut^*=0.0960 \mt$. 
\equ{ventuno} avoids large coefficients and  follows 
from the BLM optimizations of $\mut/\hat{m}_t(\mt)$ and $\mt/\hat{m}_t(\mt)$
\cite{sirqcd},
with  $a(\hat{\mut})$ and $a(\mut^*)$  evaluated 
from $a(\mz)$ using five active flavors. 
The corresponding FAC and PMS 
optimizations lead to nearly identical results \cite{sirqcd}.
In this approach,  $\mut$ is obtained from $\mt$ via
\equ{ventuno} and inserted in the various amplitudes such as 
\equ{diciannove}. The second strategy followed in the paper is based on the 
conventional $\mt$ parametrization and \equ{ventibis}.
In both approaches we  include very
small $O(\alpha\hat{\alpha}_s^2)$ effects 
arising from the light isodoublets \cite{light}
and from an inverse top mass  expansion 
of the $(t-b)$ isodoublet contribution \cite{private}.
As inputs in the Tables we employ 
$\mz=91.1863$ GeV, 
$\as(\mz)=0.118$ \cite{alphas}, $\Delta\alpha_{had}= 0.0280$ \cite{jeg},
$\mt=175$ GeV\footnote{Results for different input values can be very easily
obtained using a Mathematica package available from the authors.}.
The $O(g^4)$ contribution to ${\rm Re}\azz(\mz^2)$
in \equ{quattro} involving $({\rm Im} A_{\gamma\smallz}(\mz^2))^2$
and effects of light fermion masses in the one-loop  corrections are also
taken into account. 

Tables 1 and 2 compare the $\mw $ and $\seff$ predictions based on 
OSI (Eqs.(\ref{dieci},\ref{diciassette})), OSII (Eqs.(\ref{quindici},
\ref{diciotto})), $\msbar$ (Eqs.(\ref{due},\ref{tre},\ref{sedici})),
with and without the complete  \gd\ corrections, respectively.
Specifically, in Table 2 we neglect the terms of \gd\
in the functions $f$ and $\Delta\rho$, as well as the
contributions of the same order in the last terms of \Eqs{quindici}{diciotto}
 and the second factor of \equ{diciassette}.
In both cases, the QCD corrections are evaluated using the 
$\mut$-parametrization, explained in \equ{diciannove} et seq.,
and \equ{ventuno}. Table 3 includes the \gd\ corrections 
but evaluates the QCD corrections using the
 $\mt$-parametrization  and \equ{ventibis}.
In all cases the last column gives the values of $\sincur$ ($\msbar$ 
evaluation).

A number of striking  theoretical features are apparent:
(a) The incorporation of the irreducible \gd\ corrections greatly decreases
the scheme dependence of the predictions, as measured by
the differences encountered in OSI, OSII, and $\msbar$.
For instance, the maximum differences in Table 2 are 11 MeV in $\mw$ and
$2.1\times 10^{-4} $ in $\seff$, while in Table 1 they are reduced 
to 2 MeV and $3\times 10^{-5} $, respectively (similar very small 
differences can be observed in Table 3). 
(b) The differences between Table 2 and Table 1 are quite small in OSI.
This means that this scheme absorbs a large fraction of the \gd\
corrections in the leading contributions.
(c) Although the QCD approaches we have considered are quite different,
Tables 1 and 3 show very close results. This is due to a curious 
cancellation of screening and anti-screening effects 
in the difference between the two formulations.

Comparing the current world average $\seff=0.23165\pm 0.00024$ with 
Tables 1 and 3, and taking into account in quadrature the errors 
associated with $\Delta \alpha_{had}$ ($\pm 2.3 \times 10^{-4}$) and  
$\delta\mt=\pm6$ GeV ($\pm 1.9 \times 10^{-4}$), 
together with an estimate of missing higher order QCD corrections ($\pm
2.4 \times 10^{-5}$) \cite{sirqcd}
we obtain a determination of $\mh$. 
As the six calculations of $\seff$ in those tables
are very close, it matters relatively little which one is employed.
Averaging
the six central values and choosing the errors to cover the range of the six
calculations, we 
 find $\mh=127^{+143}_{-71}$GeV, compatible with MSSM expectations. The 
strength of this determination rests on the 
fact that unaccounted two-loop effects are not enhanced by
$(\mt^2/\mw^2)^n$ $(n=1,2)$ factors or large logarithms 
$\ln (\mz/m_f)$.
They are expected to affect \seff\ in the fifth decimal, which is consistent
 with the variations observed in 
Tables 1 and 3. Because of the approximate
exponential dependence of $\mh$ on $\seff$, this precision 
is important in order to 
obtain a reasonably accurate determination of $\mh$.
For instance, a 0.1\% difference in the theoretical calculation of \seff\
induces a $\approx$55\% shift in the \seff\ determination of $\mh$ and its
$1\sigma$ bound.
Corresponding to the above determination of $\mh$, one finds from Tables 1 
and 3 the prediction $\mw=80.367\pm0.048$GeV, in very good agreement 
with the current world average $\mw=80.356\pm 0.125$GeV.

\vspace{1.5cm}
We are grateful to  K.~Chetyrkin, G.~Ganis, W.~Hollik, and M.~Steinhauser for 
useful communications and interesting discussions. A.S.~would like to thank
the Max-Planck Institut in Munich and the Benasque Center for Physics
in Benasque, Spain, for their kind hospitality during the summer of 1996,
when part of this work was done, and M.~Passera for useful calculations 
involving the resummation formulae. His work was supported in part by NSF grant
PHY-9313781.

\section*{Appendix}
\appendix
\begin{appendletter}
In the following we adopt the notation and conventions of \efe{physlett}.
The additional contributions  to $\bar{f}$ and $\Delta\bar{\rho}$, in 
units $ N_c\, x_t/(16\pi^2)$, are given by
\begin{eqnarray}
\bar{f}_{add}^{(2)}&=&
{{10}\over 3} + {1\over {3{c^2}}} + 
  4\,{c^2}B0[\mw^2,0,\mw^2] - 
  \left( {{11}\over 3} + {1\over {3\,{c^2}}} + 4\,{c^2} \right)
 B0[\mw^2,\mw^2,\mz^2] 
\nonumber\\&&
+   {{\left( 11 - 8\,{c^2} \right) \ln {c^2}}\over 
    {6\, s^2 }} - 
  \left( {{11}\over 3} + {1\over {3\,{c^2}}} \right) \ln {\it zt},\\
\Delta\bar{\rho}^{(2)}_{add}&=&
{{542}\over {27}} - {2\over {3\,{c^2}}} - {{800\,{c^2}}\over {27}} +
  \frac1{3}\left( 1 + 26\,{c^2} + 24\,{c^4} \right) 
       B0[\mz^2,\mw^2,\mw^2] +   4\,{c^2}B0[\mw^2,0,\mw^2]
\nonumber\\&&
 -   \left( {11\over 3} + {1\over {3\,{c^2}}} + 4\,{c^2} \right) 
  B0[\mw^2,\mw^2,\mz^2] - 
  \left( {2\over 3} + {{4\,{c^2}}\over 3} - 8\,{c^4} \right)    \ln {c^2}
\nonumber\\&&
 + \left( {c^{-2}} -{{38}\over 3}  + {{34\,{c^2}}\over 3} 
 \right) \ln {\mt^2\over {{\mu^2}}}
 +   {{2\left( 3 - 62\,{c^2} + 74\,{c^4} + 36\,{c^6} \right) 
      \ln {\it zt}}\over {9\,{c^2}}}.
\end{eqnarray}

Fixing  
the one-loop result  to the $\msbar$ calculation of \efe{rel},
the \gd\ two-loop contribution to $\Delta\hat{k}^{(2)}(M_Z)$ in units 
$N_c/(16\pi^2)\, (\hat{\alpha}/(4\pi \hat{s}^2))\, M_t^2/(\ccur M_Z^2) $
reads \cite{zako}
\begin{eqnarray}
\Delta\hat{k}^{(2)}&=&
{{-211 + 24\,{\it ht} + 462\,{\scur} - 64\,{\it ht}\,{\scur
}}\over {432}} + 
  \left( {3\over 8} - {{{\scur}}\over 3} \right) 
   B0[\mz^2,\mw^2,\mw^2] - {\ccur\over6} \ln \ccur 
\nonumber\\&&
+  {{\left( ht-4 \right) {\sqrt{{\it ht}}}
      \left(  8\,{\scur}-3 \right) g(ht)}\over {108}} + 
  {{\left( 6 + 27\,{\it ht} - 10\,{{{\it ht}}^2} + {{{\it ht}}^3} \right) 
      \left(  8\,{\scur} -3\right) \ln {\it ht}}\over 
    {108\left( {\it ht}-4 \right) }} 
\nonumber\\&&
-  \left( {1\over4} + {2\over9}{\scur} \right) \ln {\mt^2\over\mu^2} + 
  {{\left(  3\,{\scur} -2\right) \ln {\it zt}}\over {18}} + 
  {{\left(  {\it ht}-1 \right) \left(  8\,{\scur} -3\right) 
      \phi ({{{\it ht}}\over 4})}\over 
    {18\left( 4 - {\it ht} \right) {\it ht}}},
\label{app1}
\end{eqnarray}
while the term to be added in the OS framework is given,
in units  $N_c x_t/(16\pi^2)$, by
\bea
\Delta \bar{k}_{add}^{(2)}=&& {{-238\,{c^2}}\over {27}} + 8{c^4} - 
   2\,{c^2}\,{\sqrt{ 4 c^2 -1}}\left( 3 + 4\,{c^2} \right) 
    \arctan ({1\over {{\sqrt{ 4\,{c^2} -1}}}}) - 
   {16\over 9}c^2\ln zt
\nonumber\\&&
+    \left( {3\over {4\,{c^2}}} - 2\,{c^2} \right) f_V(1) + 
   4\,{c^2}{ g_V}({c^{-2}}) - 7\,{c^2}\,\ln c^2 - 
   {{17\over3}}c^2 \ln {\mu^2\over \mz^2}, 
\eea
 where the functions $f_V(x)$ and $g_V(x)$ are defined in Eqs.(6d) and (6e)
of \cite{dg}.
\end{appendletter}

\newpage

\renewcommand{\arraystretch}{1.2}
\begin{table} 
\[
\begin{array}{|c|c|c|c|c|c|c|c|}\hline
& \multicolumn{3}{|c|}{\mw ({\rm GeV})} & \multicolumn{3}{|c|}{\seff} &
\scur\\\hline 
\mh  & {\rm OSI} & {\rm OSII}  & \msbar & {\rm OSI} & 
{\rm OSII} & \msbar & \msbar
 \\  \hline\hline
65 &80.404 &80.404 &80.406   &0.23132   &0.23132 &0.23130 & 0.23120 \\ \hline
100  &80.381   &80.381 &80.383 &0.23152  &0.23154 & 0.23151&0.23142  \\ \hline
300  & 80.308&80.307 &80.309   & 0.23209& 0.23212 & 0.23209& 0.23200 \\ \hline
600  & 80.255 & 80.254 &80.256 & 0.23248 &0.23250 & 0.23247& 0.23239 \\ \hline
1000 &80.216 &80.215 & 80.216 & 0.23275 &0.23277&0.23275& 0.23267\\ \hline 
\end{array}            
\]
\label{table1}
\caption{\sf 
Predicted values of  $\mw$, $\seff$, and $\sincur$.
OSI: Eqs.(\ref{dieci},\ref{diciassette});
OSII:  Eqs.(\ref{quindici},\ref{diciotto});
$\msbar$: Eqs.(\ref{due},\ref{tre},\ref{sedici}).
QCD corrections based on $\mut$-parametrization and \equ{ventuno}. 
$\mt=175$GeV, $\as(\mz)=0.118$, $\Delta\alpha_{had}=0.0280$.
}
\end{table}

\renewcommand{\arraystretch}{1.2}
\begin{table} 
\[
\begin{array}{|c|c|c|c|c|c|c|c|}\hline
& \multicolumn{3}{|c|}{\mw ({\rm GeV})} & \multicolumn{3}{|c|}{\seff} &
\scur\\\hline
\mh  & {\rm OSI} & {\rm OSII}  & \msbar & {\rm OSI} & 
{\rm OSII} & \msbar & \msbar
 \\  \hline\hline
65  &80.409 &80.419 & 80.417 &0.23132 &0.23111 &0.23124 & 0.23114 \\ \hline
100  &80.385 &80.395 &80.393 &0.23154   &0.23135 &0.23145 &0.23135  \\ \hline
300  & 80.311& 80.316& 80.318  &0.23212 &0.23203 &0.23203 & 0.23195 \\ \hline
600  &80.257 & 80.258 &80.263  &0.23251 &0.23247 & 0.23242  & 0.23235 \\ \hline
1000 & 80.215 &80.214 &80.222   &0.23279  &0.23280 &0.23271 &0.23264\\ \hline 
\end{array}            
\]
\label{table2}
\caption{\sf 
As in Table 1, but excluding \gd\ corrections (see text).
}
\end{table}
\renewcommand{\arraystretch}{1.2}
\begin{table} 
\[
\begin{array}{|c|c|c|c|c|c|c|c|}\hline
& \multicolumn{3}{|c|}{\mw ({\rm GeV})} & \multicolumn{3}{|c|}{\seff} &
\scur\\\hline
\mh  & {\rm OSI} & {\rm OSII}  & \msbar & {\rm OSI} & 
{\rm OSII} & \msbar & \msbar
 \\  \hline\hline
65  & 80.405& 80.404  & 80.406  &0.23132 &0.23134 &0.23130 &0.23121  \\ \hline
100  &80.382 &80.381 & 80.383 &0.23153 & 0.23155 &0.23152 & 0.23142 \\ \hline
300  &80.308 &80.306 & 80.308 &0.23210 &0.23214 &0.23210 & 0.23201   \\ \hline
600  & 80.254& 80.252  &80.254  &0.23249 & 0.23252 & 0.23249&0.23241  \\ \hline
1000 & 80.214 &  80.213 &80.214   &0.23277 &0.23279 &0.23277 &0.23269\\ \hline 
\end{array}            
\]
\label{table3}
\caption{\sf 
As in Table 1, but with QCD corrections based on $\mt$-parametrization
and \equ{ventibis}.
}
\end{table}

\end{document}